\documentclass[aps,prb, twocolumn,superscriptaddress,oneside,floatfix,amsmath,showpacs,amssymb]{revtex4}

\usepackage{amssymb}
\usepackage{graphicx}

\begin{document}

%\title{Superconducting critical temperature dome shape in granular aluminum \\driven by the phase stiffness}%\\
%or\\

%\title{First direct evidence of sub-gap excitations in granular aluminum\\ close to the superconductor-insulator transition}
%\title{Weak to strong coupling evolution in granular aluminum superconductivity\\close to the superconductor-insulator transition}
\title{Electrodynamics of granular aluminum from superconductor to insulator:\\observation of collective superconducting modes}

%\author{F.~L.~Emmard and G. PadeBol}
%\affiliation{Univ. Grenoble Alpes, Inst NEEL, F-38000 Grenoble, France}
%\affiliation{CNRS, Inst NEEL, F-38000 Grenoble, France}
\author{F.~Levy-Bertrand}
\email{florence.levy-bertrand@neel.cnrs.fr}
\affiliation{Univ. Grenoble Alpes, CNRS, Grenoble INP, Institut N\'eel, 38000 Grenoble, France}

\author{T.~Klein}
\affiliation{Univ. Grenoble Alpes, CNRS, Grenoble INP, Institut N\'eel, 38000 Grenoble, France}

\author{T.~Grenet}
\affiliation{Univ. Grenoble Alpes, CNRS, Grenoble INP, Institut N\'eel, 38000 Grenoble, France}

\author{O.~Dupr\'e}
\affiliation{Univ. Grenoble Alpes, CNRS, Grenoble INP, Institut N\'eel, 38000 Grenoble, France}

\author{A.~Beno\^it}
\affiliation{Univ. Grenoble Alpes, CNRS, Grenoble INP, Institut N\'eel, 38000 Grenoble, France}

\author{A.~Bideaud}
\affiliation{Univ. Grenoble Alpes, CNRS, Grenoble INP, Institut N\'eel, 38000 Grenoble, France}

\author{O.~Bourrion}
\affiliation{Laboratoire de Physique Subatomique et de Cosmologie, Universit\'e Grenoble Alpes, CNRS, 53 rue des Martyrs, 38026 Grenoble Cedex, France}

\author{M.~Calvo}
\affiliation{Univ. Grenoble Alpes, CNRS, Grenoble INP, Institut N\'eel, 38000 Grenoble, France}

\author{A.~Catalano}
\affiliation{Laboratoire de Physique Subatomique et de Cosmologie, Universit\'e Grenoble Alpes, CNRS, 53 rue des Martyrs, 38026 Grenoble Cedex, France}

\author{A.~Gomez}
\affiliation{Centro de Astrobiolog'a (CSIC-INTA), Ctra. Torrej—n-Ajalvir km.4, 28850 Torrej—n de Ardoz, Madrid, Spain}

\author{J.~Goupy}
\affiliation{Univ. Grenoble Alpes, CNRS, Grenoble INP, Institut N\'eel, 38000 Grenoble, France}

\author{L. Gr\"unhaupt}
\affiliation{Physikalisches Institut, Karlsruhe Institute of Technology, 76131 Karlsruhe, Germany}

\author{U. v. Luepke}
\affiliation{Physikalisches Institut, Karlsruhe Institute of Technology, 76131 Karlsruhe, Germany}

\author{N. Maleeva}
\affiliation{Physikalisches Institut, Karlsruhe Institute of Technology, 76131 Karlsruhe, Germany}

\author{F.~Valenti}
\affiliation{Physikalisches Institut, Karlsruhe Institute of Technology, 76131 Karlsruhe, Germany}

\author{I.~M.~Pop}
\email{ioan.pop@kit.edu}
\affiliation{Physikalisches Institut, Karlsruhe Institute of Technology, 76131 Karlsruhe, Germany}

\author{A.~Monfardini}
\email{alessandro.monfardini@neel.cnrs.fr}
\affiliation{Univ. Grenoble Alpes, CNRS, Grenoble INP, Institut N\'eel, 38000 Grenoble, France}

\begin{abstract}
We report on a detailed study of the optical response and $T_c-\rho$ phase diagram ($T_c$ being the superconducting critical temperature and $\rho$ the normal state resistivity of the film) of granular aluminum, combining transport measurements and a high resolution optical spectroscopy technique. The $T_c-\rho$ phase diagram is discussed as resulting from an interplay between the phase stiffness, the Coulomb repulsion and the superconducting gap $\Delta$. We provide a direct evidence for two different types of well resolved sub-gap absorptions, at $\omega_1\simeq\Delta$ and at $\Delta\lesssim\omega_2\lesssim2\Delta$ (decreasing with increasing resistivity).

\vspace{0.3cm}
To cite: 

F. Levy-Bertrand et al, Phys. Rev. B \textbf{99}, 094506 (2019).

DOI: 10.1103/PhysRevB.99.094506
\end{abstract}

%superconductor to insulator transition, granular aluminum, superconducting coupling strength, superconducting gap, superconducting critical temperature, nano-patterning of superconductors.

%\pacs{}
%PACS 74.	Superconductivity
%PACS 74.25.F-	Properties of superconductors Transport properties
%PACS 74.25.Bt	Properties of superconductors Thermodynamic properties
%PACS 74.25.Op	Properties of superconductors Mixed states, critical fields, and surface sheaths
%PACS 74.45.+c	Properties of superconductors Proximity effects; Andreev reflection; SN and SNS junctions

\maketitle

\section{Introduction}

In principle, superconductors are perfect mirrors and no optical absorptions are expected to be observed below twice the superconducting gap, $2\Delta$ \cite{Varma_intro} (for a fully gapped s-wave symmetry). Indeed, amplitude fluctuations of the superconducting order parameter are expected to give rise to a (scalar) mode at 2$\Delta$, decoupled from electromagnetic waves and the dispersive phase fluctuation (Goldstone) mode~\cite{Nambu,Goldstone} transforms into a non-dispersive plasma mode, well above the superconducting gap, in presence of unscreened long range Coulomb interactions~\cite{Anderson1,Anderson2}. However, an excess of optical absorption below 2$\Delta$ has been observed in various disordered superconductors. In NbN and InO, this excess has been attributed to the existence of an amplitude (Higgs) mode~\cite{Higg_InO} which could turn into a sub-gap excitation in the vicinity of a quantum critical point (while approaching a superconductor-to-insulator transition). Alternatively, in granular aluminum, this effect has been discussed as an evidence for Goldstone modes~\cite{Goldstone__Pracht}  turned into a sub-gap excitation due to the coupling of the linear dispersion with a characteristic finite momentum set in by disorder. 

We present here a detailed study of the electrodynamic properties of a series of granular aluminum thin films, combining transport measurements and sub-THz optical spectroscopy. Granular aluminum is formed of superconducting nanometric grains of pure aluminum coupled by Josephson barriers through aluminum oxide, and can be tuned from a superconductor to an insulator by varying the Josephson coupling. The  superconducting-insulator transition is reached when the aluminum oxide barriers are too large, preventing a phase coherence to develop throughout all grains~\cite{Efetov}. The superconducting critical temperature then presents a dome shape~\cite{Abeles, Cohen,Deutscher_1973,these_Bachar,Mott_Bachar, Dome_Pracht}, whose origin is still debated, with a critical temperature, $T_c$  reaching 2-3~K (depending on grain size morphology) at the maximum of the dome, which is significantly higher than the  $T_c\sim1$~K of pure aluminum. 

Our optical spectroscopy technique, combining an optical dilution fridge (with a base temperature of $\sim$100~mK) with a Martin-Puplett spectrometer of high resolution (1~GHz) from 0~GHz to 300~GHz enabled us to observe two different types of sub-gap optical absorptions. We observe a first series of well defined absorption peaks at a frequency $\omega_1$ of the order of $\Delta$ and a second larger peak at a frequency $\omega_2$ which decreases progressively from $\sim 2\Delta$ towards $\sim \Delta$ as the superconductor-insulator transition is approached. The onset of those sub-gap absorptions occurs in the vicinity of the maximum of the superconducting dome, for 100-1000~$\mu\Omega$.cm room temperature resistivity. In the same resistivity range we observed (i) a change from a positive to negative temperature dependence of the normal state resistivity, (ii) an increase of the coupling strength ratio $\Delta/T_c$ from $\sim1.78$ to  $\sim2.10$, and, (iii) we estimate that the phase stiffness $J$ falls below the geometrical Coulomb repulsion energy. Finally we show that the insulating regime is reached when $\Delta \sim J$. 

\section{Experimental}

We performed optical spectroscopy and transport measurements on ten samples from low (sample \#A) to high (sample \#J) room-temperature resistivity, spanning the phase diagram of granular aluminum from superconductor to insulator. The partial oxygen pressure was increased from samples \#A to \#J while e-beam evaporating aluminum at 0.3 nm/second on sapphire substrates held at room temperature. A study of the films structure prepared in similar conditions allows us to estimate that the grain size is of the order of $\sim$3-4~nm in our films~\cite{Deutscher_1973}. The films are 20~nm thick except sample \#H that is 30~nm thick and sample \#J that is 50~nm thick. On each film, a 3~mm$\times$0.4~mm rectangle has been lithographied in order to perform resistivity measurements.  Table \ref{table_list_sample} lists the different samples.

\begin {table}[h]
\caption {\textbf{Samples studied} where $\rho$ is the normal state resistivity, $d$ is the film thickness, $T_c$ is the critical temperature, $\Delta$ is the superconducting gap, $L_k$ is the kinetic inductance.} 
\label{table_list_sample} 
\begin{center}
\begin{tabular}{|c|c|c|c|c|c|}
\hline
&$\rho$($\mu\Omega$.cm)&d(nm)&$T_c$(K)&$\Delta$(K)&$L_k$(pH/sq)\\
\hline
\#A&40&20&1.90$\pm$0.02&3.4$\pm$0.3&20\\
\hline
\#B&80&20&2.04$\pm$0.03&3.6$\pm$0.1&33\\
\hline
\#C&160&20&2.17$\pm$0.02&3.8$\pm$0.1&80\\
\hline
\#D&220&20&2.17$\pm$0.02&3.9$\pm$0.1&110\\
\hline
\#E&900&20&2.08$\pm$0.02&4.3$\pm$0.3&325\\
\hline
\#F&1600&20&2.03$\pm$0.06&4.4$\pm$0.4&430\\
\hline
\#G&2000&20&1.99$\pm$0.06&4.3$\pm$0.3&535\\
\hline
\#H&3000&30&1.91$\pm$0.04&3.9$\pm$0.1&620\\
\hline
\#I&11400&20&&&\\
\hline
\#J&20500&50&&&\\
\hline
\end{tabular}
\end{center}
\end {table}

\begin{figure}
\begin{center}
\resizebox{8cm}{!}{\includegraphics{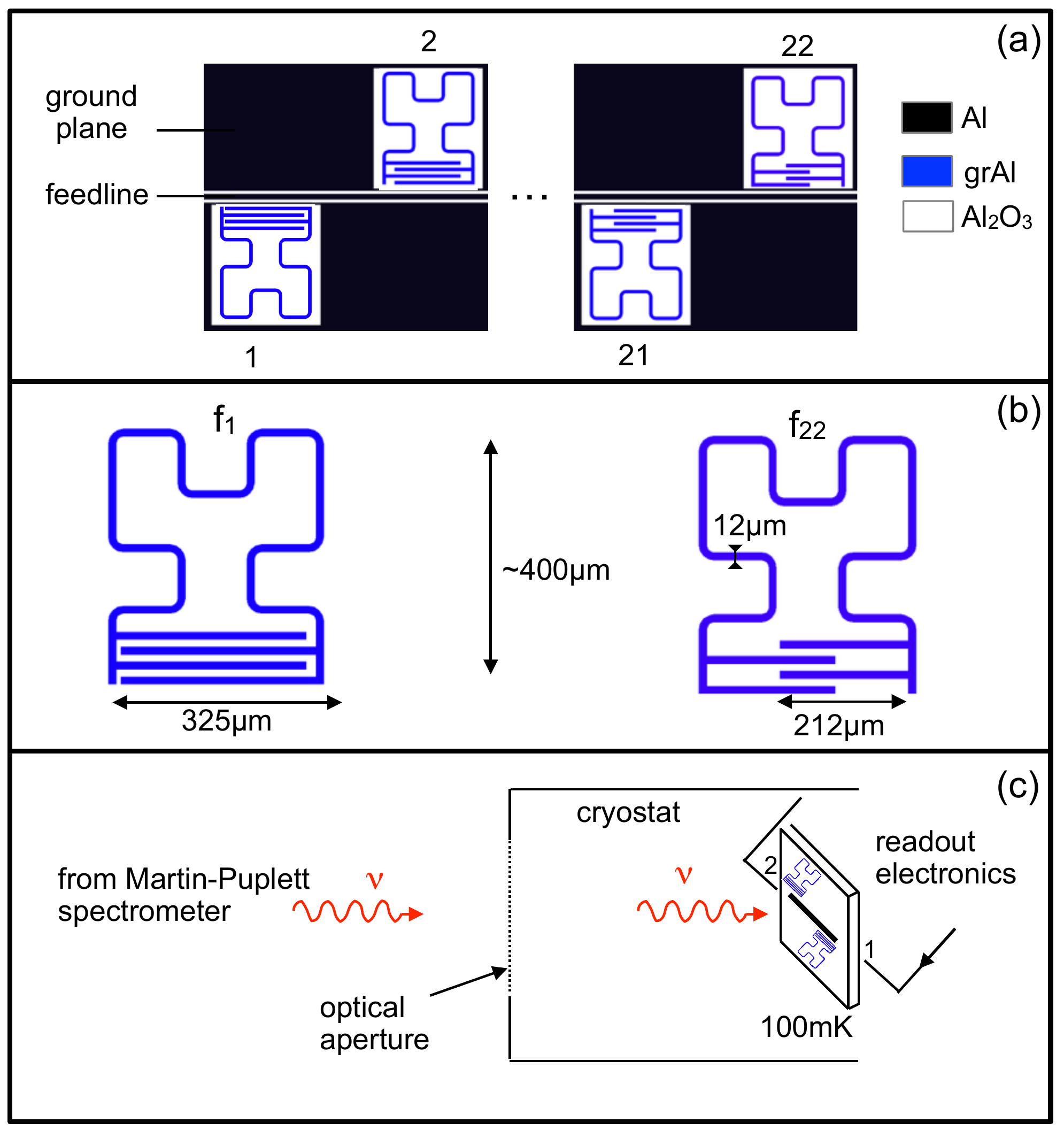}}
\caption{\textbf{Resonators design and experimental set-up.} (a) Sketch of the twenty-two resonators made out of granular aluminum, (grAl in blue)  are coupled to the common feed-line made out of pure aluminum (Al in black). The feed-line and ground-plane are in a coplanar waveguide configuration. The dielectric  employed is sapphire ($Al_2O_3$ in white). (b) Detailed design of the first ($f_1$) and last ($f_{22}$) resonators. The Hilbert fractal inductor is identical for all the resonators. The four capacitor fingers length are varied from one resonator to another: from the shortest ($f_{22}$) to the longest ($f_{1}$). (c) Experimental set-up. Incident photons coming from a Martin-Puplett spectrometer at room temperature illuminates the resonators through the optical apertures of the dilution refrigerator. The variation of the resonance frequencies are measured simultaneously for all the resonators by a special readout electronics.}
\label{fig0}
\end{center}
\end{figure}

The optical spectroscopy technique used here is inspired by millimeter astrophysics observation techniques~\cite{Day,NIKA2}. For each granular aluminum composition, from sample \#A to \#H, superconducting microwave resonators were lithographied and cooled down to 100~mK in an optical dilution fridge (those measurements could not be performed in sample  \#I and \#J for which the resistivity remains finite down to the lowest temperatures). The optical spectroscopy measurements consisted in monitoring the resonators resonance frequencies $f$ while varying the optical incident photon energy. The energy of the incident photon was spanned from 0 to 3~THz with a resolution of $\sim$1~GHz thanks to a Fourier-Transform spectrometer with a  300~K black body radiation source. An optical low pass filter in the dilution fridge limits the incident photon range to 300~GHz. Figure~\ref{fig0} presents the resonators design and a schematic view of the experimental set-up employed to perform the optical spectroscopy measurements.

The resonance frequency $f=1/(2\pi\sqrt{LC})$  is of the order of few GHz (from 2 to 6~GHz) and  is varied from one resonator to another by varying the capacitor fingers length (see panel (b) of figure~\ref{fig0}). The inductance is identical for all the resonators and can be decomposed in a geometric and a kinetic part: $L=L_g+L_k$. The shift of the resonance frequency is due to the change of the kinetic inductance that is inversely proportional to the superfluid density $n_s$~\cite{Day, Diener}: 
\begin{eqnarray}
\frac{\delta f}{f}=-\frac{\alpha}{2}\frac{\delta L_k}{L_k}=\frac{\alpha}{2}\frac{\delta n_s}{n_s}
\end{eqnarray}
where $\alpha=L_k/(L_k+L_g)$ is the kinetic inductance ratio. The superfluid density may be affected through two different mechanisms by an incident photon $h\nu$. First, when $h\nu>2\Delta$,  the incident photon may break Cooper pairs~\cite{Day}. Second, for  $h\nu<2\Delta$, superconductors are usually considered as perfect mirrors. One key point of this latter mechanism is the photon absorption. To be absorbed, the energy of an incident photon has to match the energy of a  superconducting collective mode.  When the photon is absorbed, the superfluid current density $J$ increases, decreasing the superfluid density $n_s$~\cite{De Gennes, Tinkham}, leading to an increase of the kinetic inductance $L_k$, which can be written as~\cite{De Gennes, Tinkham, Lorentz}:
\begin{eqnarray}
L_k(J)=L_k(0)[1+J^2/J_*^2]
\end{eqnarray}
where $J_*$ is proportional to the critical current $J_c$ ($J_*=2/3^{3/2}J_c$ in thin films). $J_*^2$ sets the incident power scale for which this second mechanism may be observed.  This value lowers  with the superfluid density (while approaching the superconductor-to-insulator transition for example). Some of us observed this second mechanism in amorphous indium oxide resonators where the collective modes at play were higher-order resonance (surface plasma) modes~\cite{Dupre_SKIDs}.  For each granular aluminum film from \#A to \#H the optical response displayed is an average over all the functional resonators and corresponds to a Fourier-Transform of the measurements (the individual resonators' responses  are displayed in the appendix). The amplitude of the frequency shift is uncorrected from the incoming source energy distribution and thus is given in arbitrary units.

%One key point of the latter mechanism is indeed the mechanism giving rise to the photon absorption. Indeed, for  $h\nu<2\Delta$, superconductors are usually considered as perfect mirrors and the energy of the incident photon has to match a superconducting collective mode to be absorbed. Some of us observed this second mechanism in amorphous indium oxide resonators where the collective modes at play were higher-order resonance (plasma) modes~\cite{Dupre_SKIDs}. For each granular aluminum film from \#A to \#H the optical response displayed is an average over all the functional resonators and corresponds to a Fourier-Transform of the measurements.  

%Those specifications have to be compared  to those of the optical spectroscopy techniques used in the latest studies on the subject~\cite{Higg_InO,Dome_Pracht,Goldstone__Pracht}: base temperature of 1.6~K and frequencies range starting from 60~GHz (up to 700~GHz). 

\section{Results and discussion}

\subsection{Transport measurements}

\begin{figure}
\begin{center}
\resizebox{8cm}{!}{\includegraphics{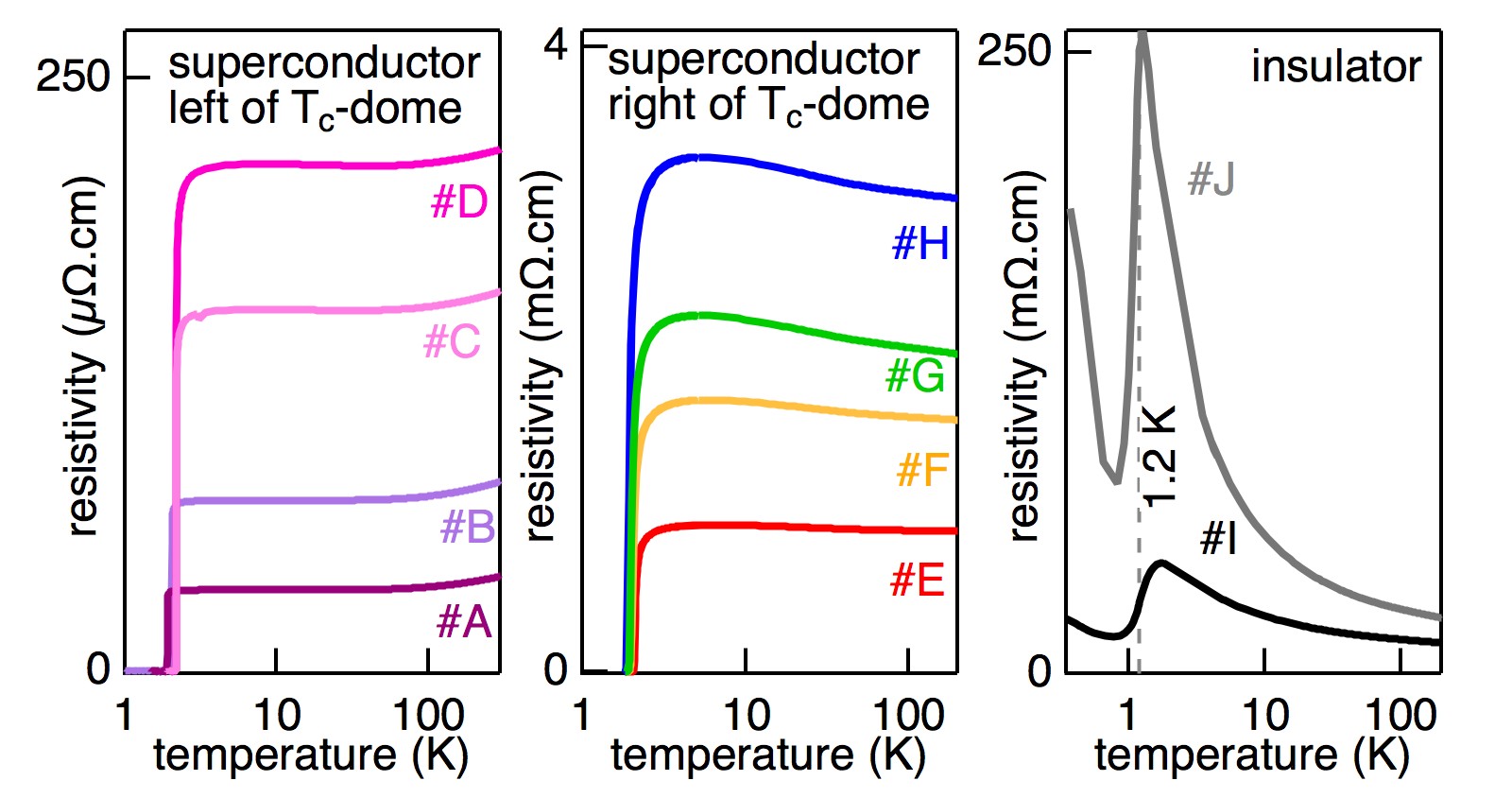}}
\caption{\textbf{Superconductor to insulator transition observed through resistivity measurements.} Resistivity as a function of temperature for ten different compositions of  granular aluminum, from less resistive, sample \#A, to  insulators, samples \#I and \#J. }
\label{fig_highR}
\end{center}
\end{figure}

\begin{figure}
\begin{center}
\resizebox{8cm}{!}{\includegraphics{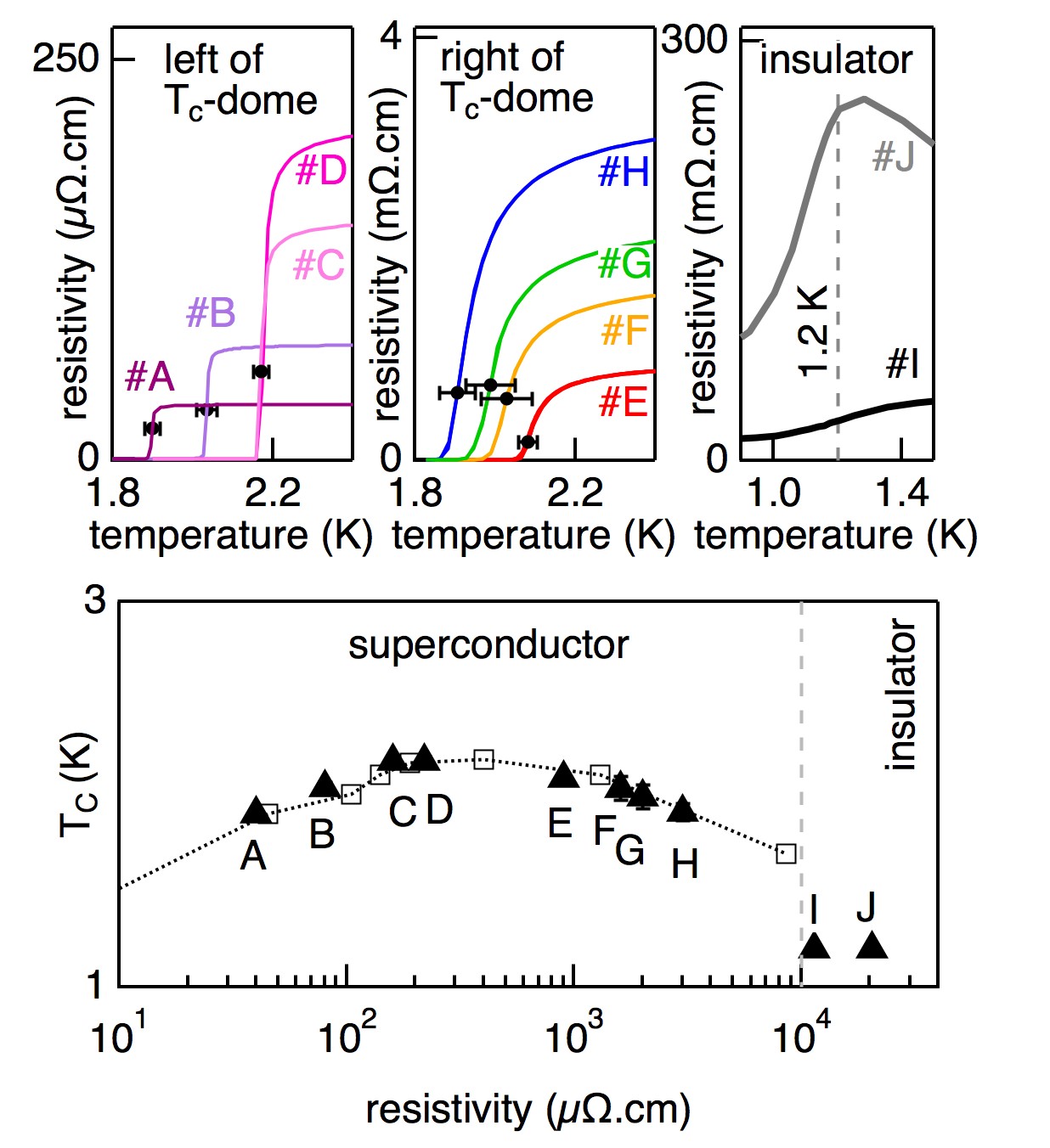}}
\caption{\textbf{Critical temperatures determined by resistivity measurements.} \textit{Top panel:} zoom on the superconducting transition of the resistivity measurements as a function of temperature. Critical temperatures, black disks, correspond to the inflection point of the resistive transitions. Error bars are the temperature intervals between zero resistivity and the inflection point.  \textit{Bottom panel:} critical temperatures versus room temperature resistivity. The dashed line connects the squares corresponding to critical temperatures of samples deposited by some of us on a different substrate (silicon) using a different e-beam deposition device.}

\label{fig_lowR}
\end{center}
\end{figure}

As shown in Figure~\ref{fig_highR}, a superconductor to insulator transition is visible on the resistivity measurements. The room temperature resistivity increases from sample \#A to sample \#J. From samples \#A to \#D, situated on the left side of the critical superconducting temperature dome, a (mainly) positive slope of the resistivity with temperature is observed  and superconductivity  develops at low temperature with $T_c$ progressively rising up to $\sim 2$K (see Figure~\ref{fig_lowR}).  As previously reported\cite{Deutscher_1973, Kondolike_Bachar, Mui, Kunchur}, a change to a negative slope of the resistivity with temperature is observed in samples \#E to \#J (right side of the critical superconducting temperature dome) but superconductivity (null resistance) still develops at low temperature in samples \#E to \#H. This change in $d\rho/dT$ may be due either to weak localisation effects~\cite{Altshuler} or to an interplay between electron-phonon coupling and disorder ~\cite{Jonson, Imry, Fratini}. 

Finally,  the temperature dependence of the resistivity in sample  \#I and \#J can be well described by a $R \propto \exp(T_0/T)^\alpha$ law (with $\alpha \sim 0.25-0.35$) both below and above the dip at 1.2K, clearly indicating that those films are insulating, at least down to 400mK.Ê As theoretically foreseen~\cite{Efetov} and previously observed~\cite{reentranceAl_Wu,reentranceAl_Jaeger}, this dip is a probable reminiscence of local superconductivity in the grains and occurs at the critical temperature of pure aluminum $T_c\sim1.2$~K (inflection point of the resistive transitions, preliminary resistivity measurements - not presented here - show that the dip disappears for increasing magnetic field, supporting this interpretation). This observation is puzzling as it is often believed that small aluminum grains may have a critical temperature larger than the bulk value~\cite{Perenboom}. The dip occurring at 1.2K could hence be the signature of the presence of big clusters of well coupled grains but it may also indicate that the enhanced critical temperature in granular aluminum requires an alternative explanation than shell effects (note that measurements on individual nanometer-scale aluminum particules indicated a constant superconducting gap value, uncorrelated with the radius of the particules~\cite{Black}).

As shown in Figure~\ref{fig_lowR}, the superconducting transitions are pretty stiff for the samples situated on the left side of the $T_c$-dome, i.e. displaying a (mainly) positive slope of the temperature dependence of the resitivity, and broaden in the samples situated on the right side of the $T_c$-dome (with a negative slope of the temperature dependence of the resistivity). The critical temperatures have then been defined as the inflection point of the resistive transitions (black disks in Fig.3). The bottom part of the figure displays the critical temperatures versus room temperature resistivity.

\subsection{Phase diagram}

\begin{figure}
\begin{center}
\resizebox{8cm}{!}{\includegraphics{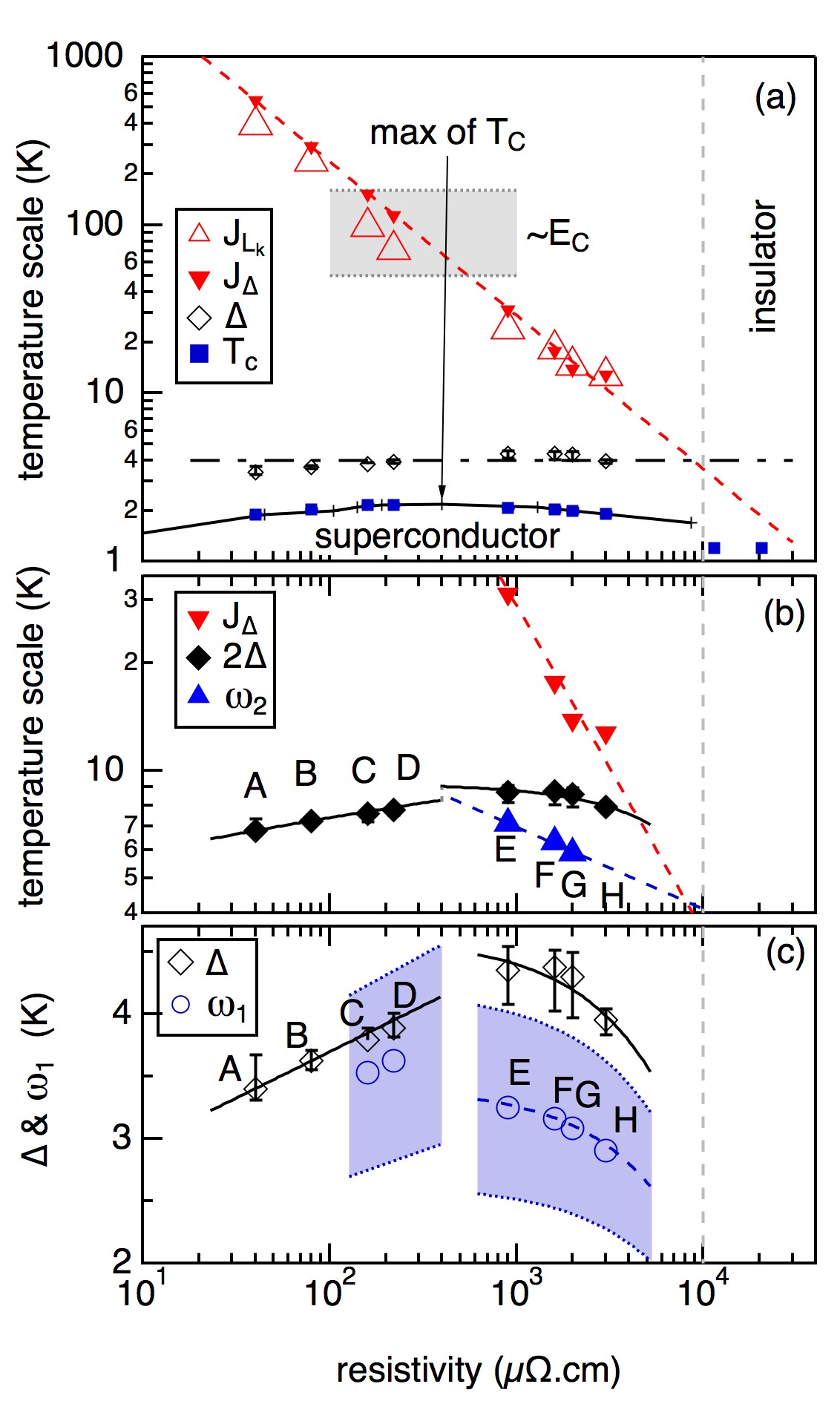}}
\caption{\textbf{Phase diagram of granular aluminum: evolution of energies at play for increasing resistivity.} \textit{Panel (a):} Critical temperature $T_c$, superconducting gap $\Delta$, phase stiffness  $J_\Delta$ and $J_{L_s}$,  and Coulomb repulsion energy $E_c$. The continuous black line with crosses corresponds to  critical temperature of samples deposited  by some of us on a different substrate (silicon). Dashed lines are linear guides to the eyes. \textit{Panel (b):} evolution of $J_\Delta$, $2\Delta$, and the sub-gap optical absorption $\omega_2$. The $\omega_2$-points correspond to the energy of the $\omega_2$-features maxima. The $\omega_2$-dashed line slope is  four times smaller than that of $J_\Delta$. \textit{Panel (c)}: evolution of $\Delta$ and $\omega_1$. The $\omega_1$-points correspond to the mid-position of the $\omega_1$-distribution. The energy distribution of $\omega_1$ is depicted by the shade area. The $\Delta$-continuous lines are guide-to-the-eyes. The $\omega_1$-dashed line is the $\Delta$-continuous line divided by 1.35.
} 
\label{fig_phasediagram}
\end{center}
\end{figure}

Figure~\ref{fig_phasediagram} displays the phase diagram of granular aluminum, presenting the evolution with resistivity of the different energies at play. Following reference~\cite{Dome_Pracht}, panel (a) displays the critical temperature, the superconducting gap (deduced from our optical measurements, see below), the phase stiffness, and the Coulomb repulsion energy. The phase stiffness of the superconducting condensate determines the phase coherence of the condensate and corresponds to the Josephson energy in the case of a network of Josephson Junctions. At zero temperature, $J$ can first be evaluated through $J_{\Delta}=\frac{\hbar}{4e^2}\frac{\pi\Delta}{R_{sq}}$ ~\cite{Efetov, these_Bachar, Dome_Pracht, Goldstone__Pracht} where $R_{sq}$ is the thin film surface resistance per square (labelled $J_\Delta$ in Figure~\ref{fig_phasediagram}). $J$ is also related to  the kinetic inductance of the superconducting resonators $L_k$ though $J_{L_k}=\frac{\hbar^2}{4e^2L_k}$ ~\cite{Rotzinger}.  By comparing their resonance frequencies, $f\sim(L_kC)^{-1/2}$, to frequencies obtained by  radio-frequency electromagnetic simulations we determined $L_k$~\cite{Dupre_SKIDs} and thus the phase stiffness labeled $J_{L_k}$. Those two $J$ values agree with each other very well. The measured data show that superconductivity is suppressed at the superconducting-insulating transition, when the phase stiffness becomes smaller than the superconducting gap. 

The Coulomb repulsion energy $E_c$, i.e the bare geometrical charging energy, that is the energy cost to transfer an electron from grain to grain, increases with the resistivity~\cite{these_Bachar}. This charging energy can be estimated for a resistivity corresponding to grains just decoupled through one unit cell of the aluminum oxide barrier by ~\cite{these_Bachar}:  $E_c=\frac{e^2}{4\pi\epsilon _0\epsilon _r d}\frac{s}{s+d/2}$ where $\epsilon _r~\sim$~8.5 is the relative dielectric constant of aluminum oxide, $s~\sim$~0.5~nm correspond to one atomic layer of the insulating barrier and $d$ is the grain size. Taking $d~\sim$~3-6~nm for one to two merging grains~\cite{Deutscher_1973}, one obtains  $E_c\sim100\pm 50$~K, which is close to the phase stiffness $J$ at the dome maximum.

Granular aluminum films may be described as disordered arrays of Josephson junctions connecting the grains. The phase diagram of granular aluminum can be interpreted as resulting from an interplay between the different energy scales that are the phase stiffness, the Coulomb repulsion and the superconducting gap~\cite{Beloborodov}. In the metallic regime, for room temperature resistivity $\rho\lesssim$100~$\mu\Omega$.cm, when superconductivity establishes, all electrons condensate and form one wave function with a unique phase. The phase is \textit{rigid} with a phase stiffness  $J$ that is orders of magnitude higher than both the Coulomb repulsion energy $E_c$ and the superconducting gap $\Delta$ (see Fig.~\ref{fig_phasediagram}a). In the vicinity of the superconducting dome maximum, for  $100\lesssim\rho\lesssim1000$~$\mu\Omega$.cm, the phase stiffness falls below the Coulomb repulsion energy. Note that the Coulomb repulsion energy $E_c$ is not a relevant energy on a microscopic scale as electron tunneling between the grains leads to a renormalization of $E_c$ down to a smaller effective Coulomb energy~\cite{Larkin,Beloborodov} $\tilde{E_c}\sim \Delta^2/J$. However, experimentally $J<E_c$ seems to coincide with the onset of sub-gap absorptions, a change from a positive to negative temperature dependence of the normal state resistivity, and an increase of the coupling strength ratio $\Delta/T_c$ from $\sim1.78$ to  $\sim2.10$. In line with reference~\cite{Emery} we suggest that for $\tilde{E_c}<J<E_c$ phase fluctuations develop leading to a decrease of $T_c$. Eventually, when $J<\tilde{E_c}$, Coulomb blockade localises the Cooper pairs within the grains, turning the system into an insulator~\cite{Efetov,Beloborodov}. The superconductor-insulator transition is theoretically expected to be reached for   $J/\tilde{E_c}\sim 1$ i.e. for $J\sim\Delta$,  which corresponds to a surface resistance per square equals to the quantum of resistance $h/(2e)^2 \sim6.4$~k$ \Omega$. Indeed, the sheet resistance of the first insulating sample is 5.7~k$\Omega$.

\subsection{Optical response and Sub-gap optical absorptions}

 \begin{figure}
\begin{center}
\resizebox{7cm}{!}{\includegraphics{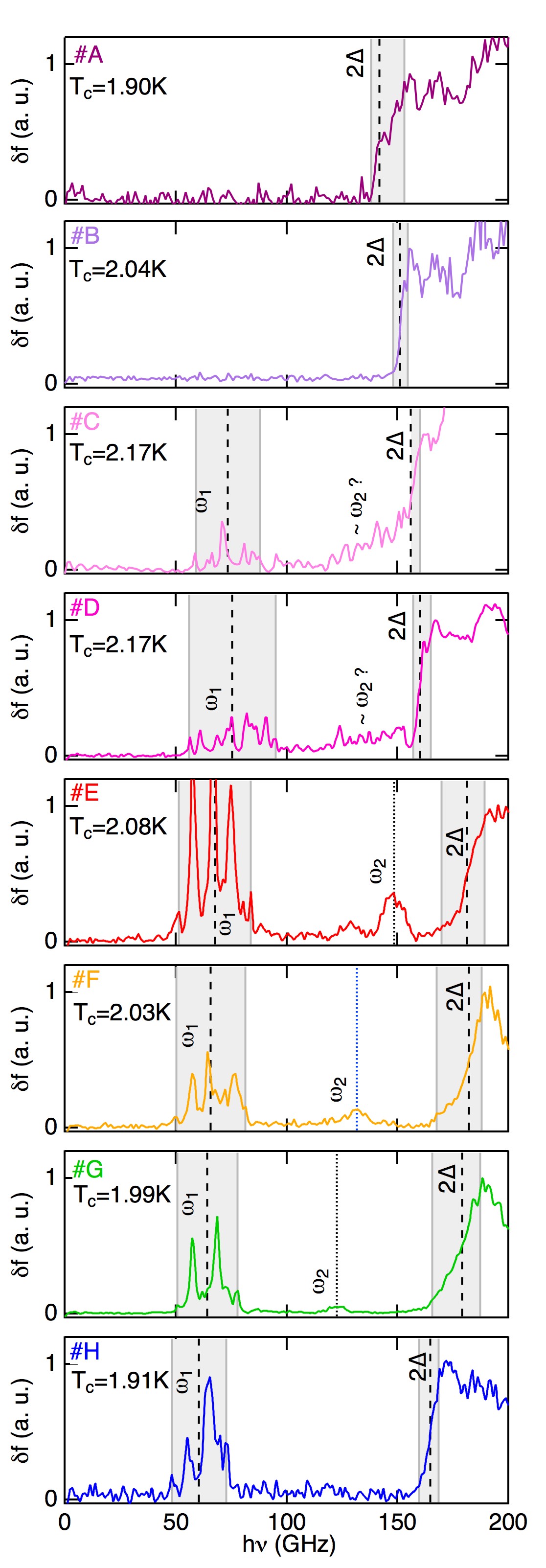}}
\caption{\textbf{Optical spectroscopy responses of superconducting granular aluminum films.}  For increasing resistivity from sample \#A to \#H, frequency shift at $\sim$100 mK as a function of the incident photon energy $h\nu$. The dash line indicates the mid-height position of the $2\Delta$-threshold, the shaded area corresponds to the 10\%-90\% height area but for sample \#C due to an excess of optical absorption just below twice the superconducting gap. Below $2\Delta$ two different types of sub-gap absorptions are observed at $\omega_1$ and at $\omega_2$.}
\label{fig_opticalresponse}
\end{center}
\end{figure}

\begin{figure}
\begin{center}
\resizebox{8cm}{!}{\includegraphics{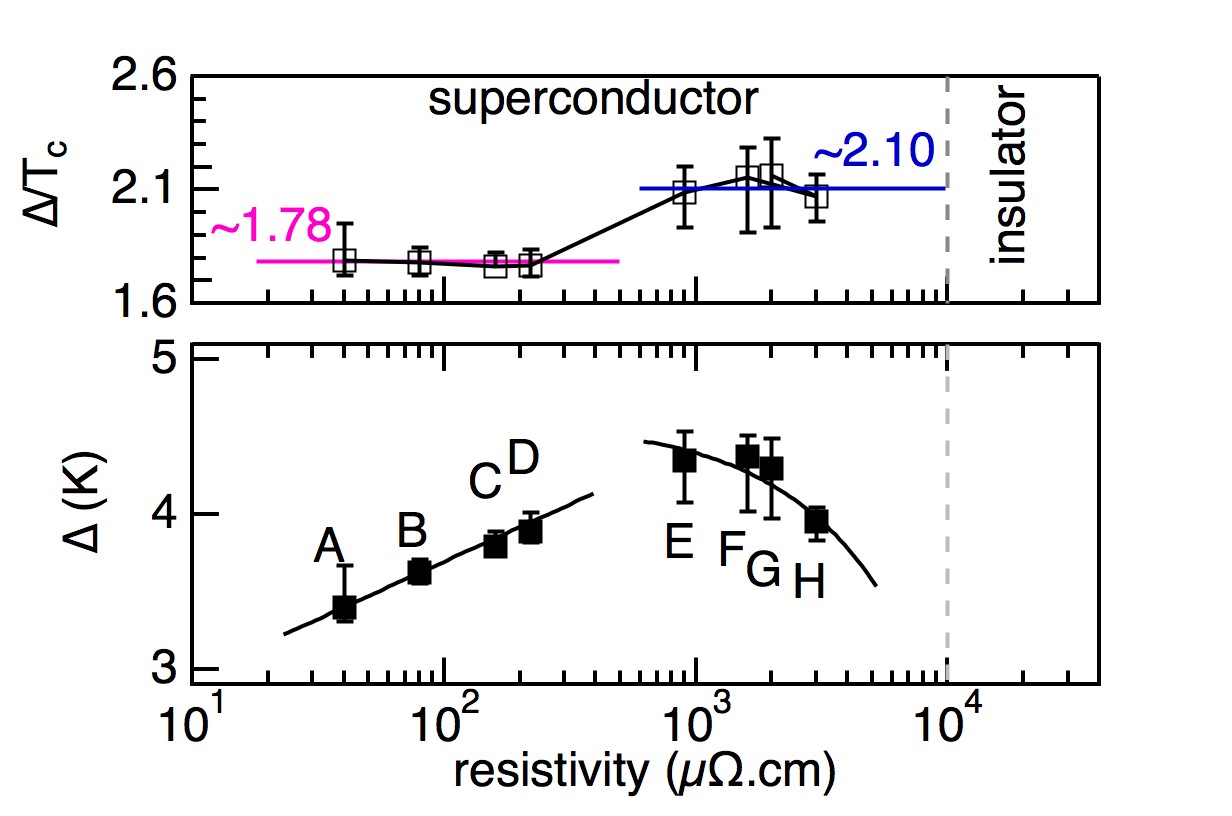}}
\caption{\textbf{Evolution with resistivity of the superconducting gap and of the ratio gap over critical temperature.}
\textit{Bottom panel:} superconducting gaps $\Delta$ measured at $\sim100$~mK as a function of  room temperature resistivity. The error bars correspond to the 10\%-90\% threshold in the optical response at $2\Delta$.  Continuous black lines are guide to the eyes.
\textit{Top panel:} coupling strength  $\Delta/T_c$ as a function of room temperature resistivity. Error bars include the error bars coming from $T_c$ (interval from inflection point to zero resistance) and the error bars coming from $\Delta$. Lines are guides to the eyes.}
\label{fig_gap_and_coupling}
\end{center}
\end{figure}

In Figure~\ref{fig_opticalresponse} we show the optical response of superconducting granular aluminum films. Twice the superconducting gap value $2\Delta$ is directly determined from the abrupt increases of the frequency shift as the photon energy $h\nu$ reaches $2\Delta$ (pair breaking).  Figure~\ref{fig_gap_and_coupling} displays the evolution of the superconducting gap and gap over critical temperature ratio as a function of the room-temperature resistivity. The top panel shows that there is a modification of the superconducting coupling parameter $\Delta/T_c$ which increases from $\sim1.78$ to  $\sim2.10$. The direct comparaison of the optical response of sample \#D ($T_c$~=~2.17~K, left side of the $T_c$-dome) to that of sample  \#E ($T_c$~=~2.08~K, right side of the $T_c$-dome) underlines that although their critical temperature are almost identical their gap are different (see  figure~\ref{fig_opticalresponse}). Note that the $\Delta/T_c$ ratio of sample  \#H, $T_c$~=~1.91~K, is clearly larger than  $\sim1.78$ including the error bars. The lines of the bottom panel intend to underline the change of the gap value.

Below $2\Delta$ the superfluid response evolves from a feature-less response for samples \#A and \#B to a response with strong sub-gap features for samples \#E to \#H. Samples \#C and \#D show an intermediate response. Our results are consistent with the excess of optical absorption observed previously below twice the superconducting gap in granular aluminum~\cite{Goldstone__Pracht} (and other disordered superconductors~\cite{Higg_InO}). This excess of optical absorption as been interpreted as a Goldstone mode~\cite{Goldstone__Pracht} of the superconducting order parameter. As we do show here that those sub-gap absorptions onset for $J\sim E_c$ and strengthen for $J<E_c$, we confirm that they are probably related to phase fluctuations of the superconducting order parameter.

 However, it is important to note that - thanks to a higher energy resolution of our spectrometer and lower working temperature - we can resolve the existence of {\it two} different types of features, occurring at $\omega_1$ and $\omega_2$ (see figure~\ref{fig_opticalresponse}). Both the energy position and  the shape of those absorption peaks are clearly different. As shown in panel (c) of figure~\ref{fig_phasediagram}, $\omega_1$ remains on the order of $\Delta$ (slightly lower) in all measured samples. This $\omega_1$-feature is rather an assembly of more-or-less distinguishable sharp features spread over few tens of GHz (the $\omega_1$ dash line indicates the mid-position of the features, and the distribution is underlined by the shaded area). In reference~\cite{Natasha} this absorption has been attributed to a two dimensional plasma phase mode. In two dimensional superconducting films plasma oscillations are expected to occur for discrete momentum values $k_n=n\times 2\pi/L$ and energy $\omega_n$ where n is an integer and L is the length of the thin film/resonator (predicted~\cite{Mishonov,Mirhashem} and observed~\cite{Buisson}). Those surface plasma modes correspond to the higher order resonance modes of a superconducting resonator and saturate at the two dimensional plasma frequency where the number of resonances then diverges~\cite{Natasha}.  In reference~\cite{Natasha} an analytical plasma dispersion established from a network of Josephson Jonction model estimates a plasma frequency of 68~GHz from the $n=1$ and $n=3$ resonances mode measured on a granular aluminium film with $\rho=4000\mu\Omega.cm$, in very good agreement with our observation. However, note that the multiple peak structure of $\omega_1$ remains unexplained within this scenario.

On the other hand, the $\omega_2$-feature is a broad response the maximum of which is indicated by a dash-line in figure~\ref{fig_opticalresponse}. This feature is only clearly distinguishable in three samples: \#E, \#F and \#G (the excess of optical response just below  $2\Delta$ in samples  \#C and \#D may be attributed to this $\omega_2$-feature), and clearly decreases from $\sim 2\Delta$ down to almost $\Delta$ as the sample resistivity increases, tending towards $J$ at the superconductor-to-insulator transition (see panel (b) of figure~\ref{fig_phasediagram}). The origin of this feature remains to be explained.

\section{Conclusion}
In conclusion, we explored the superconducting critical temperature dome shape of granular aluminum by combined state-of-the art optical spectroscopy and resistivity measurements. In the vicinity of the dome maximum, we evidenced a superconducting coupling modification from  $\Delta/T_c\sim1.78$ to  $\Delta/T_c\sim2.10$. Within the same region, we observe the occurrence of sub-gap features in the optical response, the change from a positive to negative temperature slope of the resistivity, and we estimate that the phase stiffness falls below the Coulomb energy. We evidenced two types of sub-gap excitations below twice the superconducting gap $2\Delta$ and studied their evolution with resistivity. 

\section*{Acknowledgments}
We are in debts to Lara Benfatto, R\'egis M\'elin, Beno\^it Dou\c{c}ot  and Olivier Buisson for discussions on the superconducting collective modes. 
We are very grateful to Pierre Lachkar for his help during the resistivity measurements on the  Physical Property Measurement System.
We acknowledge the contributions of the Cryogenics and Electronics groups at Institut N\'eel and LPSC. 
This work is supported by the French National Research Agency through Grant No. ANR-16-CE30-0019 ELODIS2 and by the LabEx FOCUS through Grant No. ANR-11-LABX-0013.
F.L-B acknowledges financial support from the CNRS  through a grant  \textit{D\'efi Instrumentation aux limites 2018}.
LG, UvL, NM, FV, and IMP acknowledge funding from the Alexander von Humboldt foundation in the framework of a Sofja Kovalevskaja award endowed by the German Federal Ministry of Education and Research. 
A.G acknowledges the financial support from project ESP2015-65597-C4-1-R.

\section{Appendix}

\begin{figure}
\begin{center}
\resizebox{7cm}{!}{\includegraphics{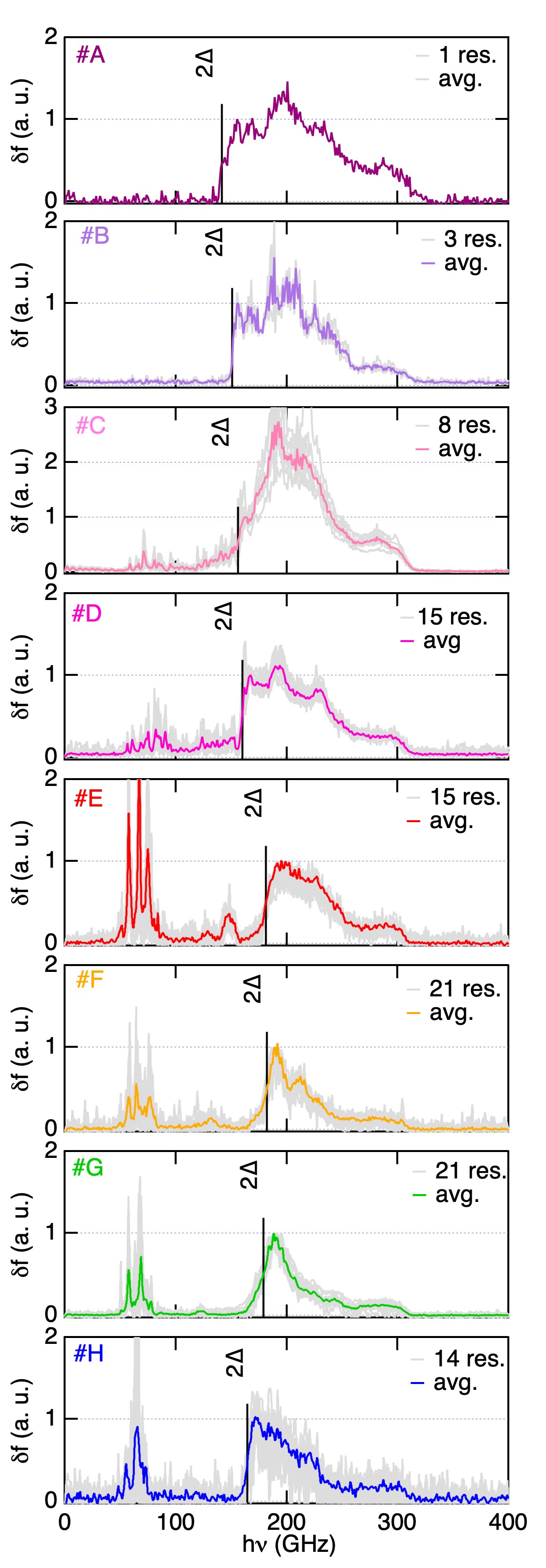}}
\caption{\textbf{Optical spectroscopy responses of superconducting granular aluminum films.}  For increasing resistivity from sample \#A to \#H,  the average (avg.) and the individual resonator (res.) frequency shift at $\sim$100 mK as a function of the incident photon energy $h\nu$. The line indicates the mid-height position of the $2\Delta$-threshold.}
\label{fig_supl}
\end{center}
\end{figure}

In Figure~\ref{fig_supl} we show, for each granular aluminum film from \#A to \#H, the average optical response and the optical response of each functional resonator. The $2\Delta$-threshold and the $\omega_2$-absorption are, within the noise, identical for every functional resonator of a given film. The $\omega_1$-features are different from resonator to resonator. An optical low pass filter in the dilution fridge limits the incident photon range to 300~GHz.

\bibliographystyle{unsrt}

\end{document}